\documentclass[aps,onecolumn]{revtex4}

\usepackage[dvips]{graphicx}
\usepackage{color}
\usepackage{amsmath}
\begin{document}

\title{{\tt {\small \begin{flushright}
{\tt ZTF-EP-14-04\\
March 2014 }
\end{flushright} } }
Comment on ``the decoupling of heavy sneutrinos in low-scale
  seesaw models''}

\author{Amon Ilakovac$^{\,a}$, Apostolos Pilaftsis$^{\,b,c}$ and Luka
  Popov$^{\,a}$\vspace{2mm}}

\affiliation{
${}^a$University of Zagreb, Department of Physics,
  Bijeni\v cka cesta 32, Zagreb,
  Croatia\vspace{1mm}\\
${}^b$CERN, Department of Physics, Theory Division, CH-1211 Geneva 23,
  Switzerland\vspace{1mm}\\
${}^c$Consortium for Fundamental Physics,
  School of Physics and Astronomy, University of Manchester,
  Manchester M13 9PL, United Kingdom
}

\begin{abstract}
\noindent
The authors of a  recent communication [arXiv:1312.5318] claim to have
traced an error in the existing literature regarding the evaluation of
the     one-loop    right-handed     sneutrino     contributions    to
lepton-flavour-violating   observables  in   supersymmetric  low-scale
seesaw  models.  In  this short  note, we  emphasize that  contrary to
those authors'  claim, our paper [arXiv:1212.5939] contains  no such a
flaw,  and  both our  analytical  and  numerical  results exhibit  the
expected  decoupling   property  of   the  heavy  sneutrinos   in  the
$Z$-penguin graphs.
\end{abstract}

\maketitle

In the  Introduction of a  recent communication~\cite{Krauss:2013gya},
the authors  are quoting our  paper~\cite{Ilakovac:2012sh}, by stating
that we have ``confirmed'' the dominance of $Z$-penguins contributions
``in most parts of parameter  space.''  However, this is an inaccurate
statement,  since we  have  only considered  \emph{one  point} in  the
mSUGRA parameter space, namely
\begin{equation}
  \label{params}
\tan\beta = 10 \,, \quad
m_0 = 1000 \textrm{ GeV} \,, \quad
A_0 = -3000 \textrm{ GeV} \,, \quad
M_{1/2} = 1000 \textrm{ GeV} \,.\nonumber
\end{equation}
These    input    parameters    are    defined   in    Eq.~(4.1)    of
\cite{Ilakovac:2012sh}  and  are  kept  fixed  in  all  our  numerical
estimates,  except when  we  were considering  the  dependence of  the
lepton-flavour-violating observables on $\tan\beta$.

\begin{figure}[!ht]
 \centering
 \includegraphics[clip,width=0.50\textwidth]{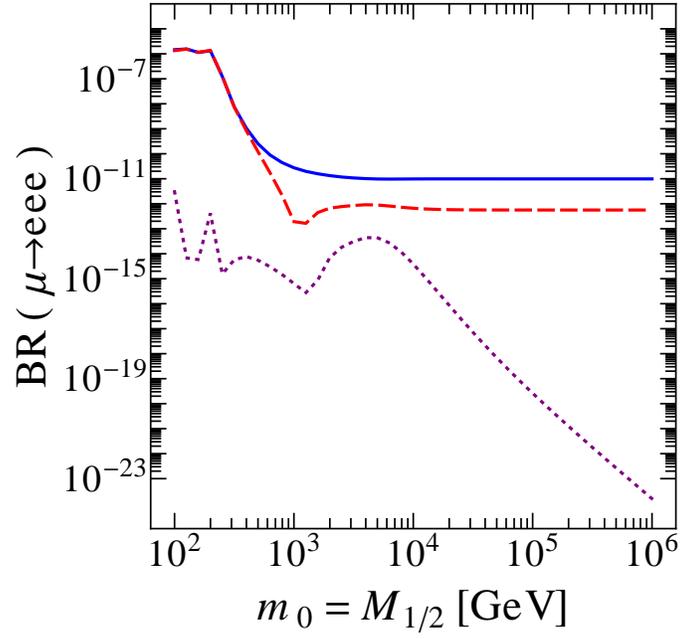}
 \caption{Numerical  estimates   of  $\textrm{BR}(\mu\to  3e)$   as  a
   function of  the mSUGRA parameter  $M_{\rm SUSY} = m_0  = M_{1/2}$,
   for  a fixed  value of  the heavy  neutrino/sneutrino mass:  $m_N =
   1$~TeV.   The (blue)  solid line  gives the  total  contribution to
   $\textrm{BR}(\mu\to  3e)$,  the  (red)  dashed line  indicates  the
   contribution  from  the  photon  penguin graph  only,  whereas  the
   (purple) dotted  line gives the  sneutrino contribution to  the $Z$
   penguin diagrams.}
\label{Fig1}
\end{figure}

The  expected  decoupling   behaviour  of  the  sneutrino  $Z$-penguin
contributions  to  $\textrm{BR}(\mu\to  3e)$  should occur  for  large
values of  $M_{\rm SUSY}$.  In our mSUGRA  framework, this corresponds
to $M_{\rm  SUSY} = m_0 =  M_{1/2}$.  We have  numerically checked the
behaviour of  the chargino-sneutrino  contribution to the  $Z$ penguin
graphs, and confirmed the existence of the decoupling for large values
of $M_{\rm SUSY}$, as shown in Fig.~\ref{Fig1}. Hence, our results are
in good agreement with those given in Fig.~1 of~\cite{Krauss:2013gya}.

\bigskip

As a  source of  the allegedly common  mistake in the  literature, the
authors  of~\cite{Krauss:2013gya}  identify this  to  be the  chargino
contributions    to     the    form    factor     $F_L$    given    in
Ref~\cite{Arganda:2005ji}.  They remark that the mistake arises due to
the incorrect use of  a relation between Passarino--Veltman integrals,
which  was  adopted  in  \cite{Arganda:2004bz}.   In  particular,  the
relation $D  C_{00} = 4  C_{00}$ should read:  $D C_{00} = 4  C_{00} -
\frac{1}{2}$.   However, in  our paper~\cite{Ilakovac:2012sh},  we did
not use  this relation, but  we have independently calculated  all the
necessary  loop  integrals,  which  are  {\em  explicitly}  listed  in
Appendix~B.  This technical mistake does not occur in our paper, as it
was further detailed in~\cite{Popov:2013xaa}.

In   conclusion,   both   our   analytical   and   numerical   results
in~\cite{Ilakovac:2012sh} exhibit the  expected decoupling property of
the  heavy sneutrinos  in  the  $Z$-penguin graphs,  and  are in  good
agreement with the results presented in~\cite{Krauss:2013gya}.

\subsection*{Acknowledgements}
\vspace{-3mm}
\noindent
The    work    of    AP     is    supported    in    part    by    the
Lancaster--Manchester--Sheffield  Consortium  for Fundamental  Physics
under STFC  grant ST/J000418/1. The  work of AI and LP is supported by
the University of Zagreb under Contract No. 202348.


\begin{thebibliography}{99}

\bibitem{Krauss:2013gya}
  M.~E.~Krauss, W.~Porod, F.~Staub, A.~Abada, A.~Vicente and C.~Weiland,
  arXiv:1312.5318 [hep-ph].

\bibitem{Ilakovac:2012sh}
  A.~Ilakovac, A.~Pilaftsis and L.~Popov,
  Phys.\ Rev.\ D {\bf 87}, no. 5, 053014 (2013)
  [arXiv:1212.5939 [hep-ph]].

\bibitem{Arganda:2005ji}
  E.~Arganda and M.~J.~Herrero,
  Phys.\ Rev.\ D {\bf 73}, 055003 (2006)
  [hep-ph/0510405].

\bibitem{Arganda:2004bz}
  E.~Arganda, A.~M.~Curiel, M.~J.~Herrero and D.~Temes,
  Phys.\ Rev.\ D {\bf 71}, 035011 (2005)
  [hep-ph/0407302].

\bibitem{Popov:2013xaa}
  L.~Popov,
  PhD thesis (Zagreb, 2013) arXiv:1312.1068 [hep-ph], pp.~73--78.

\end{thebibliography}
\end{document}